\documentclass{pasj00}              

\draft

\begin{document}
\SetRunningHead{S. Kato}{}
\Received{2008/00/00}
\Accepted{2008/00/00}

\title{Frequency Correlation of QPOs 
Based on a Resonantly-Excited Disk-Oscillation Model}

\author{Shoji \textsc{Kato}}
\affil{2-2-2 Shikanodai-Nishi, Ikoma-shi, Nara, 630-0114}
\email{kato.shoji@gmail.com, kato@kusastro.kyoto-u.ac.jp}

%


\maketitle

\begin{abstract}

In previous papers, we have proposed a model that the high-frequency 
quasi-periodic oscillations (HF QPOs) observed in black-hole and 
neutron-star X-ray binaries are inertial-acoustic oscillations that are 
resonantly excited 
on a one-armed deformed disk by nonlinear couplings between the
oscillations and the disk deformation. 
In this paper we show that in addition to the inertial-acoustic waves, 
one-armed corrugation waves are also excited in the deformed disks.
They are low-frequency oscillations.
We examine frequencies and their correlations among the inertial-acoustic 
oscillations and the corrugation waves that are excited, in order to know 
whether they can describe observed frequency
correlations among kHz QPOs and low-frequency QPOs (LF QPOs) in 
neutron-star X-ray binaries.
The results seem to well describe the observed correlations in Cir X-1, if we
adopt $M= 1.5\sim 2.0 M_\odot$ and $a_* \sim 0.8$, where $M$ is the mass of
the central star and $a_*$ is the dimensionless spin parameter of the metric.
Finally, assumptions involved in this disk-oscillation model are briefly 
summarized and discussed.
    
\end{abstract}

\section{Introduction}

In previous papers (Kato 2004, 2008), we showed that in a {\it deformed} relativistic
Keplerian disk, inertial-acoustic oscillations (and g-mode oscillations) are 
resonantly excited.
The deformation required for the excitation is a warp or a one-armed plane-symmetric
spiral pattern.

The inertial-acoustic oscillations resonantly excited in the above deformed disks  
well describe the 3:2 twin high-frequency quasi-periodic oscillations (HF QPOs)
observed in black hole X-ray binaries, if the disk region where the oscillations 
are excited is surrounded by a hot corona (torus) (Kato and Fukue 2006).
In kHz QPOs in neutron-star X-ray binaries, the frequencies of the twin QPOs 
change with time with correlation.
These correlated time change of twin kHz QPOs can be accounted for, 
if the deformation has a time-dependent precession (e.g., Kato 2007)\footnote{
In the paper by Kato (2007), mode identification of disk oscillations to
observed QPOs was not relevant.
This will be corrected in this paper.
}.
Any successful models of QPOs of neutron-star X-ray binaries, however, should 
account for the observed frequency correlations not only between twin kHz QPOs but 
also those among kHz QPOs and low-frequency QPOs (LF QPOs) 
(Psaltis et al. 1999; Boutloukos et al. 2006).

Concerning disk oscillation modes that are excited in one-armed deformed disks, 
we find by a careful inspection of mathematical stability analyses by
Kato (2008) that the one-armed c-mode oscillations are also resonantly
excited in addition of inertial-acoustic oscillations (and g-mode oscillations)
(see subsection 4.2).
This is of interest in relation to observations, since the one-armed c-mode oscillations 
have low frequencies and thus they are one of possible candidates of 
low-frequency QPOs (LF QPOs).
Disk oscillations that are excited in one-armed deformed disk are thus 
inertial-acoustic oscillations (and g-mode oscillation) and c-mode oscillations.

In this paper, we examine frequency-frequency correlations among 
the disk oscillations that are resonantly excited, in order to
compare them with observed correlations of QPOs.
The oscillation modes considered here are three oscillation modes,
i.e., two basic inertial-acoustic oscillations among excited inertial-acoustic
ones and the one-armed c-mode oscillation. 
Comparisons with observations are made to Cir X-1, since frequencies of twin kHz
QPOs and LF QPOs to this source are carefully examined recently
(Boutloukos et al. 2006).

\section{Overview of Resonantly-Excited Disk-Oscillation Model}

We present here the essential part of the model (Kato 2004, 2008), although the 
model has various variations.
The disk we consider is a geometrically thin, relativistic disk.

\subsection{Disk Deformation}

The disks are assumed to be deformed from an axially-symmetric steady 
state by some external or internal cause.
The deformation is taken to be eccentric in the disk plane, i.e., $m_{\rm d}=1$, 
where $m_{\rm d}$ is the azimuthal wavenumber of the deformation.
Concerning the symmetry of the deformation with respect to the equatorial plane,
two cases are considered.
One is the case where the deformation has no vertical asymmetry, i.e.,
$n_{\rm d}=0$, where $n_{\rm d}$ is zero or a positive integer specifying the
number of node(s) in the vertical direction.
The other is the case where the deformation is asymmetric with respect to
the equator, i.e., $n_{\rm d}=1$.
The latter deformation (i.e., $m_{\rm d}=1$ and $n_{\rm d}=1$) is 
a kind of tilt or warp.

The deformation is assumed to have a time-dependent precession, i.e.,
it rotates in the azimuthal direction with time-dependent angular velocity. 
The angular velocity of the precession is denoted by $\omega_{\rm p}$.
If $\omega_{\rm p}$ is positive the precession is prograde, while it is 
retrograde when $\omega_{\rm p}<0$.
In the case of neutron star, unlike the case of black hole, it
has a surface, and the disk surrounding the star has 
strong radiative and magnetic couplings with the star.
We assume that this is one of possible causes of time-dependent precession 
of disks.
For example, Pringle (1992, 1996) and Maloney et al. (1996) have shown that
radiative force from the central star can lead to corrugation of the disk.
Even initially planar disks are unstable to warping (Pringle 1996).
Movies made by Pringle and his collaborators show that the disk precession is 
time-dependent and it can drastically change even the direction of the precession. 

\subsection{Disk Oscillations}

We consider oscillations on the disks described above.
Disk oscillations are generally described by ($\omega$, $m$, $n$), where $\omega$
and $m$ are the angular velocity and azimuthal wavenumber ($m=0,1,2,...$)
of the oscillations, respectively, and $n$ is an integer ($n=0,1,2,...$) describing
the number of node(s) of the oscillations in the vertical direction (see Kato et al
2008 for detailed classification of disk oscillations).
For a set of ($\omega$, $m$, $n$), there are two different kinds of modes of
oscillations, except for the case of $n=0$.
In the case of $n=0$, we have inertial-acoustic oscillations (p-mode) alone,
which consist of nearly horizontal motions (motions in the disk plane).
In the case of $n\geq 1$, we have two different modes of oscillations.
One is gravity oscillations (g-mode).
The other is corrugation waves (c-mode) ($n=1$) or vertical-acoustic 
oscillations (vertical p-mode)($n\geq 2$).

The oscillation modes that are treated in this paper are the inertial-acoustic 
mode ($n=0$) [and the g-mode\footnote{
In this paper we implicitly consider the g-mode oscillations together with the inertial-acoustic
oscillations, since
they are treated in the same mathematical framework as the inertial-acoustic waves,
except when we consider their propagation region.
}
($n\geq 1$)] and the c-mode oscillations ($n=1$) with $m=1$.

\subsection{Nonlinear Resonant Processes}

A nonlinear coupling between the disk deformation characterized by  
($\omega_{\rm p}$, $m_{\rm d}$, $n_{\rm d}$) and an oscillation with
($\omega$, $m$, $n$) brings about oscillations described by
($\omega\pm \omega_{\rm p}$, $m\pm m_{\rm d}$, $n\pm n_{\rm d}$), where
various combinations of $\pm$ are possible.
These oscillations are hereafter called intermediate oscillations.
These intermediate oscillations have resonant interaction with the disk at particular
radii where the dispersion relation of the intermediate oscillations is
satisfied.
There are two kinds of resonance for the same set of ($\omega\pm \omega_{\rm p}$,
$m\pm m_{\rm d}$, $n\pm n_{\rm d}$).
In one of them, resonance occurs through horizontal motions, while it
occurs through vertical motions in the other one.
We call the former resonance a horizontal resonance, while
the latter resonance a vertical resonance (Kato 2004, 2008).

After making the resonant interaction with the disk, the intermediate oscillations
nonlinearly couple with the deformed part of the disks to feedback to the
original oscillation of ($\omega$, $m$, $n$) (see figure 1 of Kato 2004).
This nonlinear feedback processes amplify or dampen the original oscillations,
depending on the types of oscillations and of resonance.
Careful stability analyses have been done by Kato (2008) (see also Kato 2004)
in the case where the deformation has no precession, i.e., $\omega_{\rm p}=0$.

\section{Conditions of Resonance and Excitation}

Here, we outline how the results derived by Kato (2008) to the case of 
$\omega_{\rm p}=0$ are
generalized to the case of $\omega_{\rm p}\not= 0$.

\subsection{The radius of resonance}

As mentioned in the previous section, there are two kinds of resonance, i.e., 
horizontal and vertical resonances.
Hereafter, we restrict our attention only to the horizontal resonance, since
in the case of no precession, the 
resonance that can excite disk oscillations is the horizontal resonance alone
(Kato 2004, 2008).
In the case of $\omega_{\rm p}=0$, the condition of the horizontal resonance is
$[\omega-(m\pm 1)\Omega]^2-\kappa^2=0$ (Kato 2004, 2008), where
$\Omega$ is the angular velocity of disk rotation and taken to be the relativistic 
Keplerian one, $\Omega_{\rm K}$, when its numerical figure is required,
since we consider geometrically thin disks.
The symbol $\kappa$ is the epicyclic frequency.
If the deformation has a precession, the condition for horizontal resonance 
is generalized to
\begin{equation}
   [(\omega\pm\omega_{\rm p})-(m\pm 1)\Omega]^2-\kappa^2=0,
\label{2.1}
\end{equation}
where arbitrary combination of $\pm$ is allowed.
Equation (\ref{2.1}) gives resonant radii as functions of $\omega$ for a given set of
$\omega_{\rm p}$ and $m$ (see figures 1 and 2).

To find whether the resonance excites or dampens the oscillations, 
we must examine (i) the sign of work done on the oscillations at the resonant radii 
and (ii) the sign of the wave energy of the oscillations 
(Kato 2008, see also Kato 2004).
For example, if a positive work is done at a resonant radius on an oscillation of 
a positive energy, the oscillation is amplified.

\subsection{Work done on oscillations}

Analyses made by Kato (2004, 2008) show that the sign of the work done 
by resonance on oscillations at a resonant radius is governed by the sign of
$\omega-(m\pm 1)\Omega$ at the resonant radius, when the
deformation has no precession.
Let us denote resonant radii given by equation (\ref{2.1}) by
$r_{\rm resonance}$ and the work done on oscillations by the resonance 
at $r_{\rm resonance}$ by $W$.
Then, in the case where the deformation has a precession with angular
frequency $\omega_{\rm p}$, we have
\begin{equation}
   {\rm sign}(W)=-{\rm sign}[(\omega\pm\omega_{\rm p})-(m\pm 1)\Omega]_
             {\rm resonance},
\label{2.2}
\end{equation}
where sign($A$) represents the sign of $A$, and the subscript resonance denotes
the value at $r_{\rm resonance}$.

The wave energy, $E$, of oscillations depends on whether the main part of
oscillations exist inside or outside the corotation radius  defined by 
$\omega-m\Omega=0$.
If it exists inside the corotation radius, the wave energy is negative.
In the case where $E>0$, the oscillations are excited by resonance, if the 
work done on the oscillations by resonance is positive $(W>0)$.
In the case of $E<0$, on the other hand, $W<0$ excites the oscillations.

\subsection{Propagation Region of Oscillations}

For oscillations to be excited efficiently, the resonant radius must be in the
propagation region of the oscillations.
In the case of inertial-acoustic oscillations with $\omega$ and $m$, the 
propagation region is described by
\begin{equation}
   (\omega-m\Omega)^2-\kappa^2>0,
\label{2.3}
\end{equation}
while it is \begin{equation}
      (\omega-m\Omega)^2-\kappa^2<0
\label{2.4}
\end{equation}
in the case of g-mode oscillations.

In addition to the inertial-acoustic oscillations and g-mode oscillations, 
we consider in this paper the c-mode oscillations ($n=1$) with $m=1$.
Their propagation region is described by (e.g., Kato 2001, Kato et al. 
1998, 2008)
\begin{equation}
       (\omega-\Omega)^2-\Omega_\bot^2<0,
\label{2.4'}
\end{equation}
when local approximations are adopted, where $\Omega_\bot$ is the vertical 
epicyclic frequency (Kato 1990)
and slightly smaller than $\Omega(=\Omega_{\rm K}$).

\section{Description of Resonantly-Excited Oscillations on Propagation Diagram}

Descriptions in the previous section give us necessary materials for specifying
the oscillation modes that are  really excited by resonance.
We now examine this issue by using the so-called propagation diagram of
oscillations.
If an arbitrarily large precession is allowed, various oscillations
can be excited.
In order to simplify situations by excluding less realistic cases,  
we restrict our attention here only to the oscillations that can be excited even 
in the limit of no precession, 
and study how the frequencies of such resonantly-excited oscillations are 
affected by the presence of precession of deformation.
That is, we restrict our attention only to horizontal resonance.
Cases of inertial-acoustic oscillations (and g-mode
oscillations\footnote
{The resonant excitation of the g-mode oscillations can be treated 
together with that of the inertial-acoustic oscillations, as mentioned before,
except for the difference of propagation region.
Excitation of g-mode oscillations is, however, less interesting compared with
that of the inertial-acoustic oscillations, since in their propagation region the
co-rotation resonance appears except for special cases, and it dampens the 
g-mode oscillations (Kato 2003; Li et al. 2003).
Hence, hereafter, we do not discuss the g-mode oscillations.
})
and c-mode oscillations are considered separately below, since the resonant radii are
different in these cases.

\subsection{Inertial-acoustic oscillations}

Examination of excitation conditions of inertial-acoustic oscillations on the 
propagation diagram has been made in a previous paper (Kato 2008) 
in the case of no precession (see figures 1 and 2 of Kato 2008).
Here we extend it to cases where the deformation has precession.

\begin{figure}
  \begin{center}
    \FigureFile(100mm,100mm){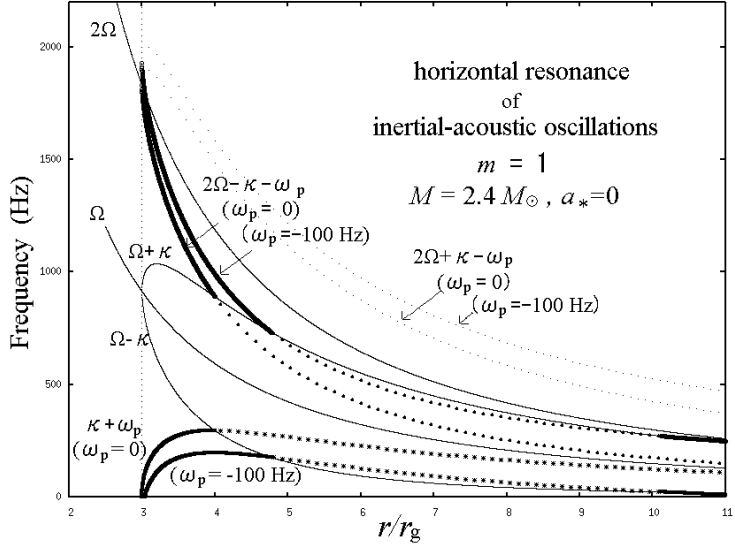}
  \end{center}
  \caption{Propagation diagram (radius-frequency relation) showing the radii
and the frequencies of inertial-acoustic oscillations (with $m=1$) that 
resonantly interact with a deformed relativistic Keplerian disk.
The disk deformation required is a warp or a one-armed plane-symmetric deformation.   
The mass of the central source is taken to be $2.4M_\odot$ with no spin.
Two case are shown where the deformation has no precession 
($\omega_{\rm p}=0$) and has
a precession of  $\omega_{\rm p}= - 100$ Hz. 
The inside area bounded by two curves labeled by $\Omega+\kappa$ and
$\Omega-\kappa$ are the evanescent region of the oscillations.
The thick curves (there are two curves for each 
$\omega_{\rm p}$, and in the case of $\omega_{\rm p}<0$ they further appear 
in the right-lower corner of the figure) 
show the radius-frequency relation of the oscillations that are excited.
The thick curves continue in the evanescent regions of oscillations.
In the evanescent region, however, excitation of the oscillations will not occur 
efficiently,
and hence the curves in the region are shown by thick but dotted curves.
At the boundaries between the propagation and evanescent regions the excitation
will occur most efficiently, since the oscillations stay there for a long time because of
vanishing of the group velocity. 
The thin dotted curves show the radius-frequency relation of resonance where
oscillations are damped.
In the case of gravity oscillations, the propagation region and the evanescent region 
are opposite to those of the inertial-acoustic oscillations.
That is, the evanescent region of the inertial-acoustic
oscillations becomes now the propagation region.
In other words, in the case of gravity oscillations, the thick dotted curves in 
figure 1 should be changed to thick curves, and the thick curves to
thick dotted ones.
}
\label{fig:figure 1}
\end{figure}

Figure 1 is the frequency-radius relation in the case where oscillations are 
one-armed (i.e., $m=1$).
The central source is taken to have no spin ($a_*=0$) and the mass $M$ 
is $2.4M_\odot$, where $a_*$ is the dimensionless spin parameter
specifying the rotation of the metric and $a_*=1$ corresponds to the extreme Kerr. 
The inner region bounded by two curves labeled by $\omega=\Omega+\kappa$ 
and $\omega=\Omega-\kappa$ is the evanescent region of the one-armed 
inertial-acoustic oscillations [cf., inequality (\ref{2.3})], and the outside of
the region is the propagation region of the oscillations.
Thick (both solid and dotted) curves give the frequency-radius relation for 
resonant excitation of one-armed ($m=1$) inertial-acoustic oscillations
[see equation (\ref{2.1}) and examine the signs of $W$ and $E$].
The dotted parts of the thick curves are, however, inside the evanescent region 
of the oscillations [see equation (\ref{2.3})], and thus excitation on the 
part will be practically inefficient.
That is, the frequency-radius relations for excitation are the thick solid curves.
Two cases of $\omega_{\rm p}=0$ and $\omega_{\rm p}=-100$ Hz are shown.
The thin dotted curves are the frequency-radius relation of resonance that
leads the oscillations to damping.

As shown in figure 1, one-armed oscillations that are excited are not unique.
In a finite range of radius, oscillations are excited with different frequencies
[see the sentence just below equation (\ref{2.1})].
Furthermore, two oscillations can be excited at the same radius.
One has higher frequencies than $\Omega$ at the resonant radius, while the other
is smaller than $\Omega$.

Here, we assume that the oscillations whose group velocity is slower will be excited 
till a larger amplitude than those with a faster group velocity, since they stay a place
for a longer time and the resonance condition is maintained for a longer time.
If we adopt local approximations, the group velocity of oscillations vanishes at the
boundary between the propagation region and the evanescent region, which is
specified by $\omega=\Omega+\kappa$ or $\omega=\Omega-\kappa$ in the 
case of one-armed oscillations.
If this consideration is adopted,  the most predominantly excited
oscillations are those with $\omega=\Omega+\kappa$ in the case of the higher
frequency oscillations and those with $\omega=\Omega-\kappa$ in the case of 
the lower frequency oscillations.
Then, combining this condition and the resonant condition given by 
equation (\ref{2.1}), we have,
as the condition determining the resonant radius,
\begin{equation}
     \kappa={1\over 2}(\Omega-\omega_{\rm p}).
\label{4.1}
\end{equation}
If the deformation has no precession, $\omega_{\rm p}=0$, this condition
gives $\kappa=\Omega/2$ and the resonance occurs at $4r_{\rm g}$ when
the central source has no spin, where $r_{\rm g}$ is the Schwarzschild radius
given by $r_{\rm g}=2GM/c^2$.
Among a series of resonant radii given by equation (\ref{2.1}), 
the resonant radius given by equation (\ref{4.1}) 
is particularly denoted hereafter 
by $r_{\rm res}$, which is a function of $a_*$, $M$, and $\omega_{\rm p}$.

If we use the notations used so far in the series of our papers, the lower-frequency
oscillation at $r_{\rm res}$ is $\omega_{\rm LL}$ and the higher-frequency
one is $\omega_{\rm H}$, i.e., 
\begin{equation}
    \omega_{\rm LL}=(\Omega-\kappa)_{\rm res}\quad{\rm and}
     \quad \omega_{\rm H}=(\Omega+\kappa)_{\rm res}.
\label{4.2}
\end{equation}

\begin{figure}
  \begin{center}
    \FigureFile(100mm,100mm){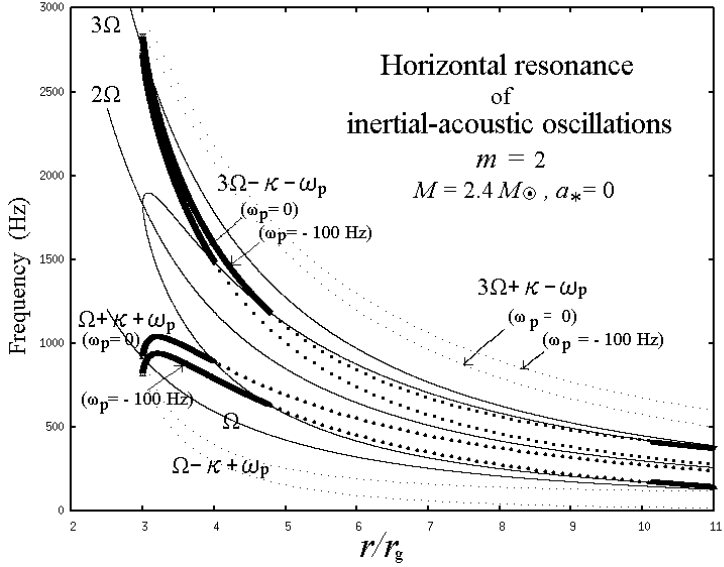}
  \end{center}
  \caption{Same as figure 1 except for $m=2$.
  In the present case of $m=2$, the evanescent region of the oscillations is
the inside region bounded by two curves of  $2\Omega+\kappa$ and
$2\Omega-\kappa$ (these curves are not labeled in the figure in order to avoid 
complexity).   }
\label{fig:figure 2}
\end{figure}

Next, we proceed to the two-armed oscillations ($m=2$).
Figure 2 is the same as figure 1 except for $m=2$.
Two cases of $\omega_{\rm p}=0$ and $\omega=-100$ Hz are again shown for 
$M=2.4M_\odot$ and $a_*=0$.
The thick solid curves and the thick dotted curves are the 
frequency-radius relations for resonant excitation of oscillations.
The thick dotted parts are, however,  inside the evanescent region of the oscillations
and thus the oscillations in that part will be practically not excited.
Thin dotted curves represent the resonance that dampens the oscillations.
As in the case of oscillations of $m=1$, the most prominently excited 
oscillations will be those in which resonance occurs at the boundary between the
propagation region and the evanescent region.
This consideration again leads to equation (\ref{4.1}).
Following again the notation used in the previous papers, we denote the frequency 
of the lower-frequency oscillations at $r_{\rm res}$ and that of the higher-frequency 
oscillations by $\omega_{\rm L}$ and $\omega_{\rm HH}$, respectively, i.e.,
\begin{equation}
    \omega_{\rm L}=(2\Omega-\kappa)_{\rm res} \quad {\rm and}
 \quad \omega_{\rm HH}=(2\Omega+\kappa)_{\rm res}.
\label{4.3}
\end{equation}

As figures 1 and 2 show, the curves of the radius-frequency relations 
for excitation of oscillations (i.e., thick solid or thick dotted curves)  cross again, 
in an outer radius, the boundary curve between the propagation region and the 
evanescent region when $\omega_{\rm p}=-100$ Hz.
In order to show this, the resonant radius, $r_{\rm res}$, obtained by solving
equation (\ref{4.1}) is shown in figure 3 as a function of $\omega_{\rm p}$
for $a_*=0.1$ and $a_*=0.3$ in the case of $M=2.0M_\odot$.
In figure 3 the $r_{\rm res}$--$\omega_{\rm p}$ relation for the corrugation
waves of $m=1$ is also shown (see the next subsection).
[In the case of corrugation waves, the resonant radius is different from $r_{\rm res}$.
That is, it is not given by 
equation (\ref{4.1}), but by equation (\ref{4.6}) (see the next subsection).]

\begin{figure}
  \begin{center}
    \FigureFile(80mm,80mm){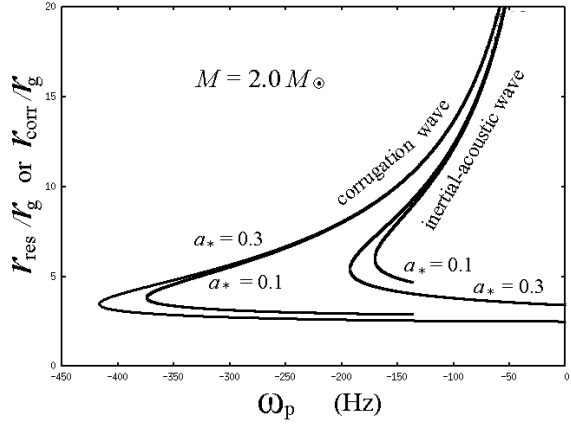}
  \end{center}
  \caption{The relation between resonant radius ($r_{\rm res}$ or $r_{\rm corr}$)
and $\omega_{\rm p}$.
 The relation is shown for inertial-acoustic oscillations and the c-mode 
oscillations with $m=1$.
Two cases of spin parameter of the central source, i.e., $a_*=0.1$ and
$a_*=0.3$,  are shown with $M=2.0M\odot$.
It is noted that there are two resonant radii for a given 
$\omega_{\rm p}$ in the case of $\omega_{\rm p}<0$.
For the case of $a_*=0$, see figure 1 by Kato and Fukue (2007).
In the case of $a_*=0$, the curve corresponding to the corrugation waves
with $m=1$ disappears in the approximation used in this paper. 
     }
\label{fig:figure 3}
\end{figure}

\subsection{C-mode Oscillations}

Here, we consider excitation of one-armed corrugation waves 
(c-mode oscillations) resulting from horizontal resonance.
The c-mode oscillations that we consider here are those of $m=1$.
They are a kind of tilt (warp) and have very low frequencies.
Their propagation region is specified by equation (\ref{2.4'}).
The boundary of the propagation region, $\omega=\Omega-\Omega_\bot$, in the
frequency-radius diagram is shown in figure 4 by a thin solid curve for two cases of
$a_*=0.3$ and $a_*=0.8$ with $M=2.0M_\odot$.
In figure 4, the left-hand side of the curve of $\omega=\Omega-\Omega_\bot$
is the propagation region of the one-armed corrugation waves.
The right-hand side of the curve is the evanescent region .
\begin{figure}
  \begin{center}
    \FigureFile(100mm,100mm){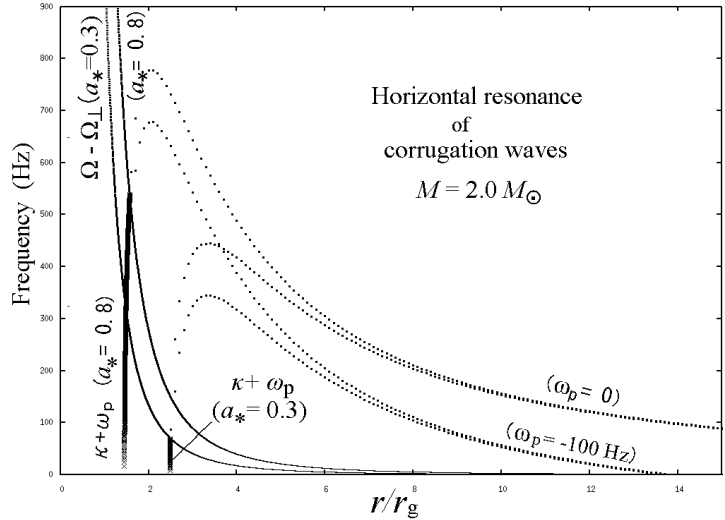}
  \end{center}
  \caption{Propagation diagram showing the radius -- frequency relation
for resonantly excited corrugation waves with $m=1$.
Two cases of  $a_*=0.3$ and $a_*=0.8$ are shown for two cases of
precession, $\omega_{\rm p}=0$ and -100 Hz, with $M=2.0M\odot$.
The curves labelled by $\Omega-\Omega_\bot$ are the boundary between 
the propagation region and the evanescent region.
The left-hand side of the curves are the propagation region.
As in figures 1 and 2, the thick solid curves represent the radius -- frequency 
relation where oscillations are excited.
Although not shown in this figure, the thick solid curves appear again in the
outer region of disks when $\omega_{\rm p}=-100$ Hz (see figure 5).
     }
\label{fig:figure 4}
\end{figure}

The horizontal resonance occurs at the radii where 
$[(\omega\pm\omega_{\rm p})-(m\pm 1)\Omega]^2-\kappa^2=0$ is satisfied
as mentioned before [see equation (\ref{2.1})].
Examination of the sign of work done on the one-armed ($m=1$) corrugation waves
at the resonant radii given by equation (\ref{2.1}) and the sign of the wave 
energy of the waves shows that among resonances specified by equation (\ref{2.1}) 
the resonance given by $\omega=\kappa+\omega_{\rm p}$ excites the  oscillations.
In figure 4, the curve of $\omega=\kappa+\omega_{\rm p}$ is denoted by
a thick solid curve when it is in the propagation region ($\omega<\Omega-\Omega\bot$)
of the oscillations, while
by a dotted curve when it is in the evanescent region.
Although two cases of $\omega_{\rm p}=0$ and $\omega_{\rm p}=-100$ Hz
are shown in figure 4, the difference of two curves of 
$\omega=\kappa+\omega_{\rm p}$ in two cases of $\omega_{\rm p}=0$
and $\omega_{\rm p}=-100$ Hz is little, 
when the curves are in the propagation region of the oscillations.
In the right-lower corner of figure 4, the curve of $\omega=\kappa+\omega_{\rm p}$
crosses again the curve of $\omega=\Omega-\Omega_\bot$ when $\omega_{\rm p}<0$, and
the curve of $\omega=\kappa+\omega_{\rm p}$ enters into the propagation region of the
corrugation waves for large value of $r$.
In order to show this, an enlargement of the right-lower corner of figure 4
is shown in figure 5.

As in the case of inertial-acoustic oscillations, the group velocity of the 
c-mode oscillations vanishes at the boundary of the propagation and 
evanescent regions.
Hence, we consider that the c-mode oscillations predominantly excited 
by the resonance are those that satisfy simultaneously $\omega=\Omega-\Omega_\bot$ 
and $\omega=\kappa+\omega_{\rm p}$.
Combination of these two equations gives
\begin{equation}
    \kappa=(\Omega-\Omega_\bot)-\omega_{\rm p}.
\label{4.6}
\end{equation}
This is the resonant condition for the most predominantly excited c-mode oscillations.
The resonant radius determined by this condition is denoted by
$r_{\rm corr}$.
The frequency of the corrugation waves at this resonant radius   
is denoted by $\omega_{\rm corr}$, i.e.,
\begin{equation}
     \omega_{\rm corr}=(\Omega-\Omega_\bot)_{\rm corr}.
\label{4.7}
\end{equation} 
Equation (\ref{4.6}) is satisfied at
two radii when $\omega_{\rm p}<0$ as is shown in figures 4 and 5.
The radius-precession relation specified by equation (\ref{4.6}) is shown 
in figure 3 in the case of $a_*=0.1$ and 0.3 with $M=2.0M_\odot$.

\begin{figure}
  \begin{center}
    \FigureFile(80mm,80mm){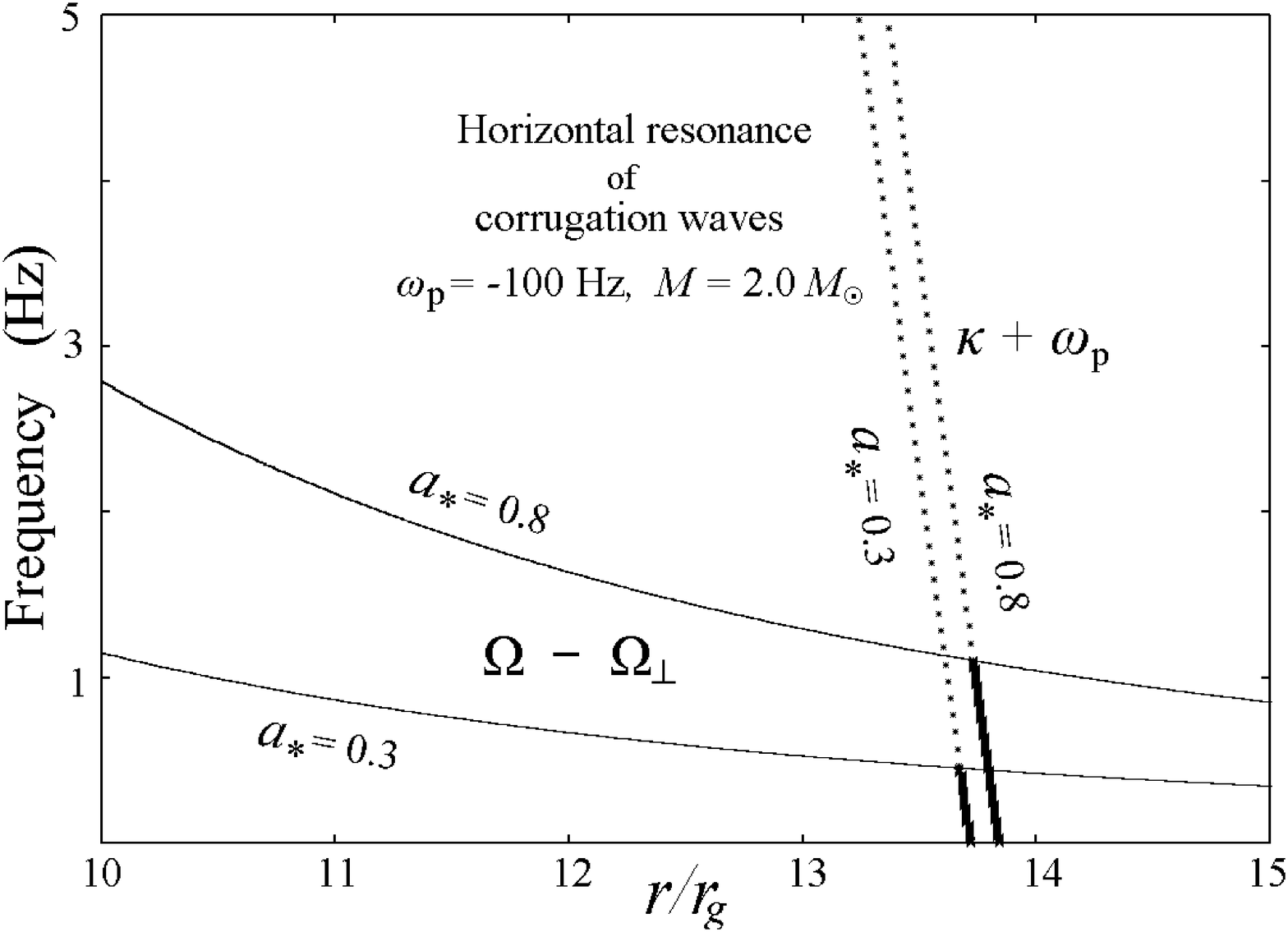}
  \end{center}
  \caption{Enlargement of the lower-right corner of figure 4.
  Notice that in the case of $\omega_{\rm p}=-100$ Hz, the dotted curve
  crosses the curve of $\omega=\Omega-\Omega_\bot$.
  Outside the crossing point,
  the dotted curve changes into a thick solid curve (i.e., resonance excites
  oscillations).
     }
\label{fig:figure 5}
\end{figure}

\section{Frequency Correlations and Comparison with Observations}

Arguments in subsection 4.1 show that the inertial-acoustic oscillations 
that are dominantly excited on the deformed disks are those whose frequencies
are specified by $\omega_{\rm LL}$, $\omega_{\rm L}$, and $\omega_{\rm H}$
[see equations (\ref{4.2}) and (\ref{4.3})].
(Oscillations with $\omega_{\rm HH}$ are not considered here, since their 
frequency is rather high and will be outside of observations.) 
Among these three oscillations, the oscillation of $\omega_{\rm H}$ will
be less important in observational view points.
This is because the propagation region of the oscillation is 
outside the resonant radius $r_{\rm res}$ and extended infinitely.
This suggests that the oscillation will not remain for a long time around 
the resonant radius, since it propagates away outside.
On the other hand, the propagation regions of waves with $\omega_{\rm LL}$ 
and $\omega_{\rm L}$ are inside the resonant radius.
The waves are, thus, partially 
trapped between  the resonant radius, $r_{\rm res}$, and the inner edge of 
the disk, and remain 
in that region for a long time, different from the wave with $\omega_{\rm H}$.

Next, we should emphasize here that 
one-armed oscillations with $\omega_{\rm LL}$ 
are not always observed only with their own frequency $\omega_{\rm LL}$.
They may be observed with the two-fold frequency when the disk region
where the oscillations are excited is surrounded by a hot corona and
when we observe high energy photons that are reprocessed in the corona
(see figures 2 -- 4 by Kato and Fukue 2006).
Based on these two considerations, we think that the frequencies related to 
the twin HF QPOs observed in black-hole and neutron-star binaries are
$2\omega_{\rm LL}$ and $\omega_{\rm L}$.

The corrugation waves that are excited are also one-armed, and thus
by the same reason mentioned above they will be observed with the 
twofold frequency, $2\omega_{\rm corr}$.
It is further noted that they are trapped inside the resonant radius, $r_{\rm corr}$.

In summary, we take the picture that the main frequencies to be observed are
\begin{equation}
     2\omega_{\rm corr}, \quad 2\omega_{\rm LL}, \quad \omega_{\rm L},
\end{equation}
in the order of increase of frequency.
As discussed below, $2\omega_{\rm corr}$ will correspond to the low-frequency QPO
(LF QPO), and $2\omega_{\rm LL}$ and $\omega_{\rm L}$ correspond to the lower and
upper kHz QPOs, respectively.

If the deformation has no precession, the frequency ratio of $2\omega_{\rm LL}$ 
to $\omega_{\rm L}(=\omega_{\rm H})$  is just 2 to 3, independent of $M$ 
and $a_*$.
We think that the twin QPOs in black-hole X-ray binaries represent this
case (Kato and Fukue 2006).
 
\begin{figure}
  \begin{center}
    \FigureFile(80mm,80mm){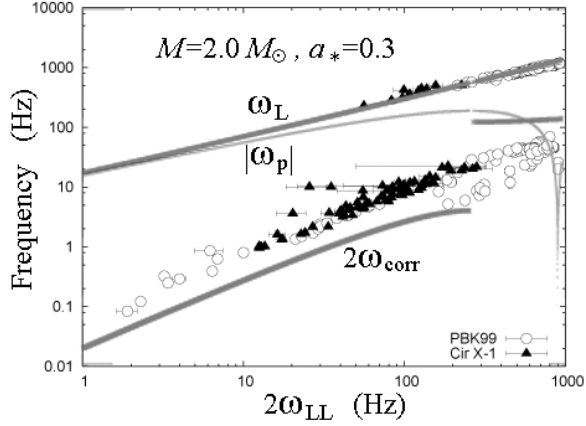}
  \end{center}
  \caption{Diagram showing the $\omega_{\rm L}$ -- $2\omega_{\rm LL}$ and 
  $2\omega_{\rm corr}$ -- $2\omega_{\rm LL}$ relations in the case of
  $M=2.0M_\odot$ and $a_*=0.3$.
  The curve of the $\vert\omega_{\rm p}\vert$ -- $2\omega_{\rm LL}$ relation
  is also shown by thin curve.
  Observational data showing the correlation between the upper and lower
  kHz QPOs and the correlation between the low frequency QPO and the lower
  kHz QPO (Boutloukos et al. 2006) are superposed.
  The filled triangles are the data of Cir X-1.
  The open circles are data of other sources of Psaltis et al. (1999), 
  which have been plotted on Boutloukos et al.'s figure. 
     }
\label{fig:figure 6}
\end{figure}

\begin{figure}
  \begin{center}
    \FigureFile(140mm,140mm){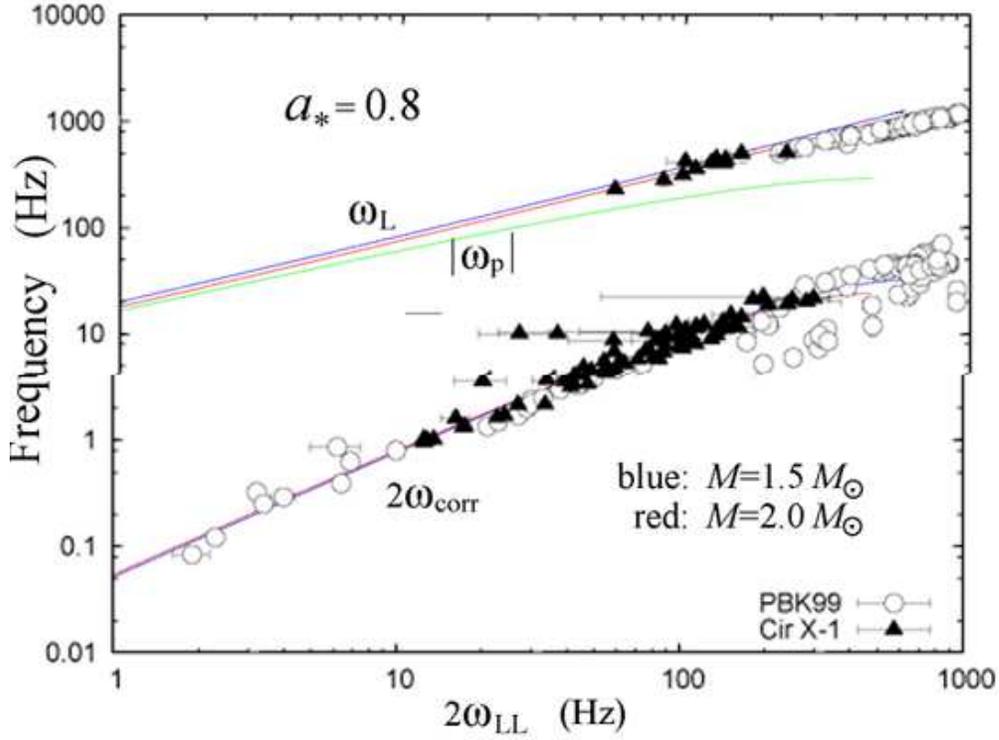}
  \end{center}
  \caption{The same as figure 6, except that two cases of ($M=1.5M_\odot$, $a_*=0.8$)
  and ($M=2.0M_\odot$, $a_*=0.8$) are drawn on the same diagram.
  The blue curves are for ($M=1.5M_\odot$, $a_*=0.8$) and the red ones are for
  ($M=2.0M_\odot$, $a_*=0.8$).
  The green curve represents the precession frequency.
  The differences between the two curves of $2\omega_{\rm corr}$ 
  are a few, except in the high frequency region.
  }
\label{fig:figure 7}
\end{figure}


In the case of neutron-star X-ray binaries, we suppose that the disk deformation
has a time-dependent precession as mentioned before (see subsection 2.1).
The frequencies $2\omega_{\rm corr}$, $2\omega_{\rm LL}$, and $\omega_{\rm L}$
are thus functions of $\omega_{\rm p}$ as well as $M$ and $a_*$.
Then, rearranging the functional relations, we obtain
$2\omega_{\rm corr}$, $\omega_{\rm L}$,  and $\vert \omega_{\rm p} \vert$ as 
functions of $2\omega_{\rm LL}$ with parameters $a_*$ and $M$.
As a typical example, these relations are shown in figure 6 for
$M=2.0M_\odot$ and $a_*=0.3$.
It is noted that this figure is drawn for $\omega_{\rm p}<0$.
When $\omega_{\rm p}<0$, we have two resonant radii for a given $\omega_{\rm p}$ (see
figure 3), i.e., an inner resonance and an outer resonance.
The curves drawn in figure 6 (and in subsequent figures) are for the frequencies 
of the outer resonance\footnote{
The inner resonance has higher frequencies and is inadequate to describe
the QPO frequencies observed in Cir X-1.
In some sources, kHz QPOs have been observed with higher frequencies, 
compared with in Cir X-1.
Such high frequency QPOs will come from the inner resonance, 
as discussed in the next section.
}.
On figure 6 the observational data of Cir X-1 (figure 9 of Boutloukos et al.
2006) are superposed, assuming that the frequency of the lower kHz QPO 
corresponds to $2\omega_{\rm LL}$.
The frequency correlations shown in figure 6 seem to qualitatively account for 
the correlations among observed QPOs of Cir X-1, but there are systematic deviations,
especially in the correlation between the $2\omega_{\rm corr}$ -- $2\omega_{\rm LL}$ 
relation and the observed LF QPO -- lower kHz QPO relation.
We have two adjustable parameters, i.e., $M$ and $a_*$.
The $2\omega_{\rm corr}$ -- $2\omega_{\rm LL}$ relation depends strongly
on the value of $a_*$, since $\omega_{\rm corr}$ is related to a
small difference between two large quantities of $\Omega$ and $\Omega_\bot$, 
and $\Omega_\bot$ becomes close to $\Omega$ with decrease of $a_*$.
On the other hand, the $\omega_{\rm L}$ -- $2\omega_{\rm LL}$ relation depends only
weakly both on $a_*$ and $M$.
The observed data of Cir X-1 seem to be well described, if $M=1.5\sim 2.0M_\odot$ and 
$a_*=0.8$ are adopted, as shown in figure 7.
It is noted here that if we adopt $a_*=0.6$ with $M=1.5\sim 2.0 M_\odot$, 
the $2\omega_{\rm corr}$ -- $2\omega_{\rm LL}$ relation runs slightly below the
observed LF QPO -- lower kHz QPO relation.

\begin{figure}
  \begin{center}
    \FigureFile(80mm,80mm){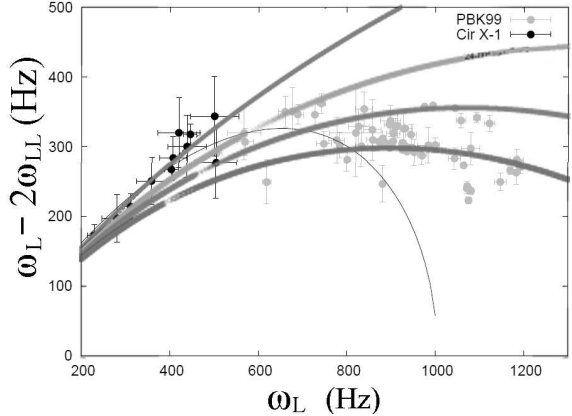}
  \end{center}
  \caption{Diagram showing the $\Delta\omega-\omega_{\rm L}$ relation, where
  $\Delta\omega$ is the difference between $\omega_{\rm L}$ and $2\omega_{\rm LL}$,
  i.e., $\Delta\omega=\omega_{\rm L}-2\omega_{\rm LL}=\kappa_{\rm res}$.
  Parameters ($M$, $a_*$) adopted are, from the uppermost to 
  lowermost curves, (2.0, 0.8), (2.0, 0.3), (2.0, 0.0),
  and (2.4, 0.0).
  The diagram showing the difference between the upper and lower kHz QPO 
  frequencies versus the upper kHz QPO frequency (Boutloukos et al. 2006)
  are superposed, assuming that
  the upper kHz QPO corresponds to $\omega_{\rm L}$.
  The black circles (they are on the left-hand side of this figure)
  are the data of Cir X-1 (Boutloukos et al. 2006) and the dark circles are 
  the data of other sources (Psaltis et al. 1999). 
  The thin solid curve is the relativistic precession model by Stella and
  Vietri (1999) (see Boutloukos et al. 2007).
  }
\label{fig:figure 9}
\end{figure}
 
\begin{figure}
  \begin{center}
    \FigureFile(80mm,80mm){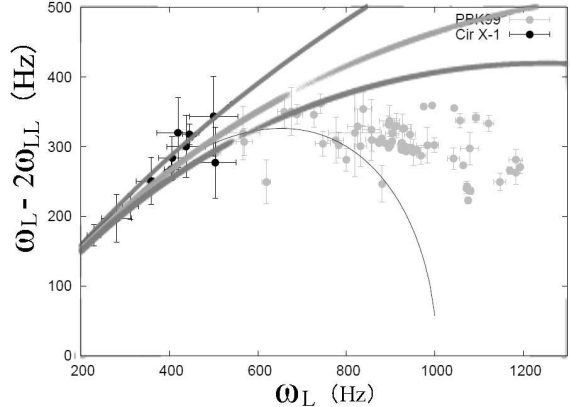}
  \end{center}
  \caption{The same as figure 8 except that $M=1.7M_\odot$ and
  $a_*=0.8$, 0.3, 0.0 from upper to lower.
  }
\label{fig:figure 9}
\end{figure}

In order to see more closely in what parameter cases the 
$\omega_{\rm L}$ -- $2\omega_{\rm LL}$ relation fits the observations,
the difference, $\Delta\omega$, between $\omega_{\rm L}$ and $2\omega_{\rm LL}$,
i.e., $\Delta\omega\equiv\omega_{\rm L}-2\omega_{\rm LL}=\kappa_{\rm res}$, is
plotted against $\omega_{\rm L}$ in figure 8 for $M=2.0M_\odot$ with some values
of $a_*$.
For a comparison, the case of $M=2.4M_\odot$ and $a_*=0$ is also shown in figure 8.
The observed frequency difference between the upper and lower kHz QPOs versus 
the upper kHz QPOs in Cir X-1 is given in figure 11 of Boutloukos et al. (2006).
The figure of Boutloukos et al. (2006) has been overlapped on figure 8.
As another example of comparison, the cases of $M=1.7M_\odot$ are shown in figure 9 
for three cases of $a_*=0.8$, 0.3, and 0.0.
Figures 8 and 9 show that a rather high value of $a_*$, say $a_*\sim 0.8$, 
is necessary in the case of Cir X-1 in order to account for the correlation between
the twin kHz QPOs.
Requirement of such a high value of $a_*$ is compatible with what required 
in the previous paragraph to describe the LF QPO -- lower kHz QPO correlation.

\section{Discussion}

We have applied a resonantly-excited disk-oscillation model to Cir X-1 in section 5.
A comparison of observational data with the model suggests that the spin of the central source
of Cir X-1 is rather high, i.e., $a_*\sim 0.8$.
Let us roughly estimate the spin frequency $\nu_{\rm s}$ of Cir X-1 by using 
the estimated value of $a_*$.
If a neutron star is rotating uniformly inside the star with $\nu_{\rm s}$,
the specific angular momentum, $a$, of the star is given by $a=2\pi \nu_{\rm s}I/cM$,
where $I$ is the moment of inertia of the star.
Since $a_*$ is related to $a$ by $a_*=(c^2/GM)a$, we have
$\nu_{\rm s}=a_*(GM^2/2\pi cI)$, which gives
\begin{equation}
    \nu_{\rm s} \sim 1.4\times 10^3\biggr({M_0^2\over I_{45}}\biggr)a_* \quad ({\rm Hz}),
\end{equation}
where $M_0=M/M_\odot$ and $I_{45}=I/10^{45}$.
If we take $M_0=2$, $M_0/I_{45}=1$ (e.g., Stella and Vietri 1999), and $a_*=0.8$,
we have $\nu_{\rm s}\sim 2.2\times 10^3$ Hz.
Such a high spin, as well as a high eccentric orbit of Cir X-1, may
suggest that the system is young.

Comparison of observations and the model suggests that the disk of Cir X-1 has a 
retrograde precession and it
varies in the range of 100 Hz $\sim$ 200 Hz (see figures 6 and 7).
For kHz QPOs to be excited by retrograde precession, the absolute value of
the precession must be less than a few hundred Hz (see figure 3).
Otherwise, inertial-acoustic oscillations cannot have resonant interaction 
with disk deformation,  i.e., we have no kHz QPOs.
That is, the range of variation of QPO frequencies is related to the 
magnitude of precession and the range of its variation.

Here, we more closely discuss the relations among precession, resonant radius, 
and frequency of resonant oscillations.
If precession is retrograde and its absolute value is
smaller than a critical value, resonance occurs at two different radii for each of 
inertial-acoustic and corrugation waves (see figure 3). 
For each case of inertial-acoustic and corrugation waves, we call
the resonance at the outer radius an outer resonance, and the resonance 
at the inner radius an inner resonance (see section 5).
In the case of Cir X-1, as shown in section 5, the resonance that
can describe observations is the outer one (see also that the gradients of
three curves of 
$\vert\omega_{\rm p}\vert$, $\omega_{\rm L}$, and  $2\omega_{\rm corr}$ 
are all positive in figures 6 and 7).
In Cir X-1, the frequencies of oscillations resonantly excited at the inner 
resonance are higher than those observed, as long as conventional mass
of neutron stars are adopted. 
The fact that oscillations at the inner resonance are not observed in Cir X-1
suggests that the inner disk region of Cir X-1 is highly disturbed.
One of possible causes of such disturbances will be
strong couplings of the inner disk region with the surface of the central source
through  strong magnetic fields and high spin of the central source.

The cases where the observed twin kHz QPOs are the oscillations excited 
at the outer resonance will be rare, compared with the cases where they are
those excited at the inner resonance, since the twin QPOs in many sources have 
higher frequencies than those in Cir X-1.
That is, in the sources where the high-frequency twin QPOs have really 
frequencies close to or higher than one kHz, they will be resonant oscillations 
at the inner resonance and the precession will be small or prograde.
It is noted that in the case where the precession is slow or prograde,
the outer resonance will be observationally less interesting, since the frequencies
excited there are too low or the outer resonance itself is absent
(see figure 3).

Let us briefly demonstrate an example in which QPOs occur at the inner resonance.
Sco X-1 has twin kHz QPOs in the 800 -- 1100 kHz range, a $\sim$45 Hz 
horizontal-branch oscillation (HBO), and a $\sim$ 6 Hz normal-branch oscillation (NBO).
Here, we examine the possibility that the observed twin kHz QPOs correspond to 
$2\omega_{\rm LL}$ and $\omega_{\rm L}$, and the 45 Hz oscillation to $2\omega_{\rm corr}$,
as in the case of Cir X-1.
The observed frequencies and their changes of the lower and upper kHz QPOs
can be roughly described by assuming that the precession of the disk deformation is 
prograde and $M\sim 2.4 M_\odot$, when $a_*$ is taken to be zero 
(see figure 3 of Kato 2007).
If $a_*$ is zero, however, there is no corrugation wave ($\omega_{\rm corr}=0$)
and we cannot account for a $\sim$ 45 Hz oscillation by our present model.
As mentioned in section 5, the frequency correlation between $2\omega_{\rm LL}$ and
$\omega_{\rm L}$ is little affected by changing the value of $a_*$, while 
the frequency of $2\omega_{\rm corr}$ is strongly affected by $a_*$.
Hence, by adjusting the value of $a_*$ we can describe a 45 Hz QPO by 
$2\omega_{\rm corr}$.
Our preliminary results show that both frequencies of twin kHz QPOs and of HBO can
be accounted for by our model with $M\sim 2.6 M_\odot$\footnote
{This value of $M$ is rather high compared with the conventional one adopted as 
the mass of neutron stars, but will be still in the range allowed theoretically and 
observationally.
} 
and $a_* \sim 0.15$,
the precession being nearly zero or weakly prograde.
A $\sim$ 6Hz normal-branch oscillation (NBO) might be a manifestation of the
slowly precessing disk deformation required in our model.

It is noted here that we have adopted $2\omega_{\rm corr}$ (not $\omega_{\rm corr}$) 
and $2\omega_{\rm LL}$ (not $\omega_{\rm LL}$) as frequencies of LF QPO and the
lower kHz QPO, respectively.
This is, however, not always the case in all sources, unless the disks are
surrounded by coronae.
Hence, some caution will be necessary when we make comparison of the model with
observations.
For example, the frequency of LF QPO may be $\omega_{\rm corr}$ in some objects.

Hereafter, we mention the basic assumptions involved in the model, and
discuss validity of these assumptions.
First, we have assumed that (i) the disk is deformed in an eccentric form
($m_{\rm d}=1$) in the state when the QPOs are observed.
There seems no clear observational evidence for or against this assumption.
There are, however, numerical simulations that support a relation between disk 
deformation and appearance of QPOs.
Magnetohydrodynamical simulations of accretion flows by Machida and
Matsumoto (2008) show a formation of a torus in the central region of disks and
its quasi-periodic deformation into a one-armed form at a certain phase of the flows.
They show that this deformation gives rise to a low-frequency quasi-periodic
variation of mass accretion rate, and further that in this deformed state, 
high-frequency QPOs 
of the order of one hundred Hz appear (the mass is taken to be 10 $M_\odot$).
It is not clear, however, whether such disk deformations are related to the disk 
deformation required in the present disk-oscillation model.

The second requirement involved in the present model is that
(ii) the deformation must have a time-dependent precession.
This time-dependent precession is necessary to bring about a change of resonant 
radius and thus to account for the frequency
change of QPOs.
It is uncertain whether such precession really exists in neutron-star binary
sources.
By couplings through magnetic and radiation fields, the accretion 
disks surrounding a neutron star will have strong interaction with the surface
of the star.
This may be one of possible causes of time-dependent precession of deformed disks
(see subsection 2.1).  
The Papaloizou -- Pringle instability in a torus, however, will not be at least the cause of
the disk precession in Cir X-1, since the precession required to describe the observations is 
retrograde.
The hectohertz QPOs might be a manifestation of the precession, but if so,
it is not clear why they are not always observed in the phases where QPOs appear.

Third, it is assumed that 
(iii) one-armed ($m=1$) disk oscillations excited on disks are observed
with the two-fold frequency.
That is, we assume that the resonant region of the geometrically thin disks 
where QPOs are excited is surrounded by a hot corona (a torus)
and the observed QPO photons are those reprocessed in the corona to higher
energy.
If such a picture is taken, we can expect the two-fold frequency in the case of 
one-armed oscillations (see figures 2 -- 4 by Kato and Fukue 2006).
At the phase when the QPOs are observed in black
hole sources (i.e., the very high state), the spectra really show coexistence of a disk
component and a corona (a torus).
It is uncertain, however, that such corona (torus) exists even in the case of 
neutron-star X-ray sources, since in the case of neutron stars, a hot corona will 
be weakened by presence of abundant soft photons from the surface of the star.

Fourth, an important assumption involved in our model is that (iv)
dominantly excited
oscillations are those with zero group velocity.
This is an approximate procedure.
In the future, it will be necessary to solve global disk oscillations with relevant
boundary conditions in order to clarify whether this approximate procedure is 
allowed as the first approximation.
Another important issue to be reminded here is whether the outward
propagation oscillations, characterized by $\omega_{\rm H}$ and
$\omega_{\rm HH}$, can be really regarded as less important,
compared with the oscillations characterized by $\omega_{\rm LL}$
and $\omega_{\rm L}$.

Unlike the epicyclic resonant model by Abramowicz and Klu{\' z}niak (2001) and
Klu{\' z}niak and Abramowicz (2001), in the present disk-oscillation model, 
the appearance of the twin QPOs is not a result of mutual interaction of
twin oscillations.
In the present model, twin oscillations are independently excited by interaction
with the disk deformation.
In this reason, there is no direct amplitude relation between the twin oscillations
in the present model.
T{\"o}r{\"o}k (2008) found a frequency -- amplitude relation of the observed twin QPOs.
Such correlations cannot be accounted for in this model, unless nonlinear
evolutions of the excited oscillations are considered.

Finally, we should notice a close relation between our present model and
the relativistic precession model by Stella and Vietri (1999).
In the latter model, the upper and lower kHz QPOs are interpreted, respectively,
as a frequency of disk rotation, $\Omega$, and periastron precession 
frequency, $\Omega-\kappa$, at a certain radius, which is a free parameter.
In the present disk-oscillation model, the frequency of the upper kHz QPO 
is $2\Omega-\kappa$ (which is close to $\Omega$), and that of
the lower one is $2(\Omega-\kappa)$.
In the present mode, the radius specifying these frequencies is the radius of
resonance, $r_{\rm res}$, which is determined by coupling between the oscillations
and the disk deformation. 
Further, in the model by Stella and Vietri (1999), the frequency of 
low-frequency QPO is the nodal precession frequency, $\Omega-\Omega_{\bot}$,
at a radius, which is again a free parameter.
In our model the frequency of low-frequency QPO (LF QPO) is 
the frequency of corrugation waves, which is $\Omega-\Omega_\bot$, but the 
radius specifying the value of $\Omega-\Omega_\bot$ is the
resonant radius given by $r_{\rm corr}$.

\bigskip

The author thanks the referee for invaluable comments on comparison of the
model with observations.
The author also thanks M. Abramowicz, W. Klu{\'z}niak, R. Matsumoto, S. Mineshige,
M. Bursa,  J. Horak, M. Machida, G. T{\"o}r{\" o}k, for helpfu discussions
during the YITP workshop YITP-W-07-14 on "Quasi-Periodic Oscillations and 
Time Variabilities of Accretion Flows".
This workshop was financially supported by the Yukawa Institute of Theoretical 
Physics at Kyoto University.

\bigskip
\leftskip=20pt
\parindent=-20pt
\par
{\bf References}
\par
Abramowicz, M. A., \& Klu{\' z}niak, W. 2001, A\&A, 374, L19 \par
Boutloukos, S., van der Klis, M., Altamirano, Klein-Wolt, M., Wijnands, R., 
     Jonker, P.G., \& Fender, R.P.  2006, ApJ, 653, 1435\par 
Boutloukos, S., van der Klis, M., Altamirano, D., Klein-Woly, M., Wijnands, 2007,
     WSPC - Proceedings, astro-ph/0701660v2 \par
Kato, S. 1990, PASJ, 42, 99 \par
Kato, S. 2001, PASJ, 53, 1\par 
Kato, S. 2003, PASJ, 55, 257 \par
Kato, S. 2004, PASJ, 56, 905\par
Kato, S. 2007, PASJ, 59, 451 \par
Kato, S. 2008, PASJ, 60 in press (arXiv: 0709.2467) \par
Kato, S., Fukue, J., \& Mineshige, S. 1998, Black-Hole Accretion Disks 
  (Kyoto: Kyoto University Press)\par
Kato, S., Fukue, J., \& Mineshige, S. 2008, Black-Hole Accretion Disks 
  -- Toward a New paradigm -- (Kyoto: Kyoto University Press)\par
Kato, S. and Fukue, J. 2006, PASJ, 58, 909 \par  
Klu{\' z}niak, W., \& Abramowicz, M. 2001, Acta Phys. Pol. B32, 3605   \par
Li, L.-X., Goodman, J., Narayan, R. 2003, ApJ, 593, 980 \par
Machida, M. and Matsumoto, R. 2008, PASJ submitted \par
Maloney, P.R., Begelman, M.C., \& Pringle, J.E., ApJ, 472, 582\par
Pringle, J.E. 1992, MNRAS, 258, 811 \par
Pringle, J.E. 1996, MNRAS, 281, 357 \par
Psaltis, D., Belloni, T., and van der Klis M.,1999, ApJ, 520, 262 \par
Stella, L., and Vietri, M. 1999, Phys. Rev. L82, 17 \par 
T{\"o}r{\"o}k, G., 2007, private communication \par

\end{document}